\documentclass[11pt,twoside]{article}

\usepackage{asp2006}
\usepackage{epsf}
\usepackage{psfig}
\usepackage{lscape}

\begin{document}
\title{Dynamic Range Improvement of GMRT Low Frequency Images}
                %%% Fill in title
\author{Peeyush Prasad, C.R. Subrahmanya}
                %%% Fill in author names
\affil{Raman Research Institute, Bangalore, India}
                 %%% Fill in author affiliations

\begin{abstract} %%% Abstract to run on from here.

This paper outlines some new observational and data processing techniques
for enhancing the dynamic range of low frequency images obtained with
the Giant Metrewave Radio Telescope.  We illustrate new software tools
developed to facilitate visibility editing and calibration as well as
other preprocessing required to enhance the dynamic range of images from
a planned survey.

\end{abstract}

%%% MAIN BODY OF TEXT GOES HERE. CONSULT "INSTRUCTIONS FOR AUTHORS USING
%%% LATEX2E MARKUP", SECTIONS 2.3-2.6 FOR HELP WITH EQUATIONS, FIGURES,
%%% AND TABLES.

\section{Introduction}   %%% Top level section head (remove "%" symbol)

The measured system parameters of the individual elements of the Giant
Metrewave Radio Telescope (GMRT) \citep{sw1991} indicate that the rms 
sensitivity achieved for GMRT images for a full synthesis at the lowest 
frequencies are only comparable to those theoretically expected for a 10-minute
snapshot observation \citep{GMRTSPECS}. In this paper, we address some of the
possible ways of improving the situation and give some preliminary
results from useful software tools being developed for visibility
editing and preprocessing.

Firstly, we note that the traditional integration times of several seconds
for each visibility point are a potential source of systematic biases in
the measured visibilities in the presence of bursty interference and fast
phase variations expected from ionospheric effects at low frequencies.
For this, we suggest taking advantage of the short term integration of
0.13 second permitted by the GMRT correlator hardware. Secondly,
by planning observations in terms of independent observing sessions
(different days) with identical antenna pointing and sidereal time range,
identification and editing of corrupted data becomes possible by examining
the cross-correlation of visibilities corresponding to identical sidereal
times  for each baseline. In other words, we advocate splitting the overall 
integration time on a field into identical observing sessions spread over 
multiple days.  
  Finally, we note that forthcoming geosynchronous navigation satellites 
\citep{ki2003} provide an interesting co-location of facility for
a continuous measurement of ionospheric delay variations along the line
of sight of the satellite from the array.  The best exploitation of this
can be made with transit observations with the GMRT, as indicated
by the novel dual-frequency mode described in Section 2.

In order to support transit observations and fast sampling of
visibilities, some minor modifications were carried out on the data
acquisition programs at the GMRT.  More importantly, the large volume 
of data generated, as well as the need for nonstandard data preprocessing
prompted us to develop new tools for facilitating visualization,
automatic editing and calibration suitable for the observing techniques
suggested by us.  Some examples from the use of these tools are presented
in Section 3.

\section{Trial Observations}

In order to carry out field trials of our observing methodology, we used the 
GMRT in transit mode over an identical LST range on three days.  The specific 
choice of antenna pointing was made to ensure that a geosynchronous satellite 
(Inmarsat 4F1) broadcasting L-band navigation signals was always within the 
field of view. The navigation signals were test signals centered at 1176 MHz 
being broadcast by the Indian Space Research Organization (ISRO) as part of a
Wide Area Augmentation Service (WAAS) \citep{ki2003} trial run.  Since the 
satellite signals were strong enough to be picked up directly by the L band 
feed of GMRT without requiring the dish collecting area, three of the GMRT 
antennas were operated in a special dual frequency mode to simultaneously 
receive 235 MHz and 1.17 GHz in the two IF bands. This was possible since the 
L band feed of the GMRT looks at the sky when the 235 MHz feed is facing the 
dish; thus enabling the 1176 MHz signal to be picked directly from the feed
pointing at the sky while the low frequency signal was received by the
feed facing the dish.  For our observations, the GMRT antennas were
nominally pointed towards the direction of the satellite in the middle
of every one-hour observing session.

The use of the GMRT correlator simultaneously for the satellite and the
celestial sky led to a conflicting requirement for fringe stopping. While
fringe stopping was necessary for the celestial signal to prevent loss
of correlation due to fringe winding, the absence of fringing for a
geosynchronous satellite would result in decorrelation of the satellite
signal due to substantial phase winding if fringe rotation was activated.
This conundrum was managed by a partial fringe stopping achieved by fooling the
correlator software such that fringe winding was contained within a radian
in the worst case for both the satellite and the celestial signal within
the 0.13 second integration time.

The arrival time differences of a geosynchronous satellite at a small
inclination ($2.4^o$) are expected to vary very slowly (dominantly
diurnal sinusoidally), while the measured phase differences(Fig. 1)
indicated small short term variations superposed on a slow drift.
Such deviations from a smooth curve are due to the variation of differential
delay caused by the ionosphere.  These can be used to provide a direct
estimate of the variation of arrival time difference caused
by the ionosphere. This in turn can be used to provide a reasonable
estimate of the phase variations at 235 MHz.  The smooth curve by
itself is useful for improving orbit estimates of the satellite, details
of which will be published elsewhere.  

  From the point of view of low frequency imaging from the GMRT, it is 
significant that the scope of
the Indian Satellite Navigation program has now been been widened  by
ISRO which plans to deploy upto 7 satellites with navigation payload in the 
near future, in addition to the first satellite due for launch during middle of 
2009.  This will give continuous and direct measurements of phase distortions 
along many lines of sight from the GMRT array. At any (low) frequency of 
observation, we can either dedicate a few GMRT antenna for such a signal 
acquisition, or, for the special case of 235 MHz observations,the dual frequency
 mode as described by us can be utilized.

\begin{figure}
\plottwo{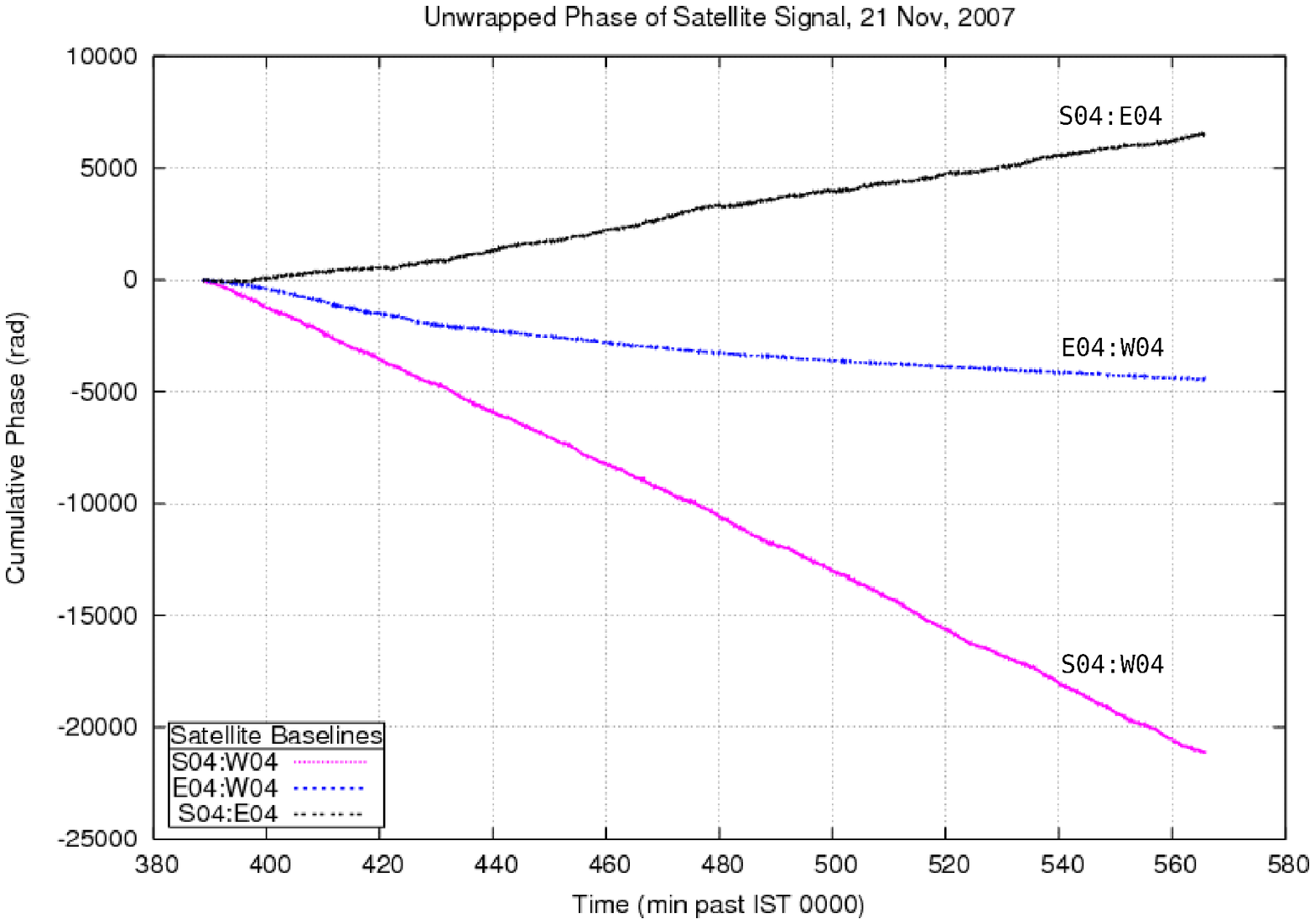}{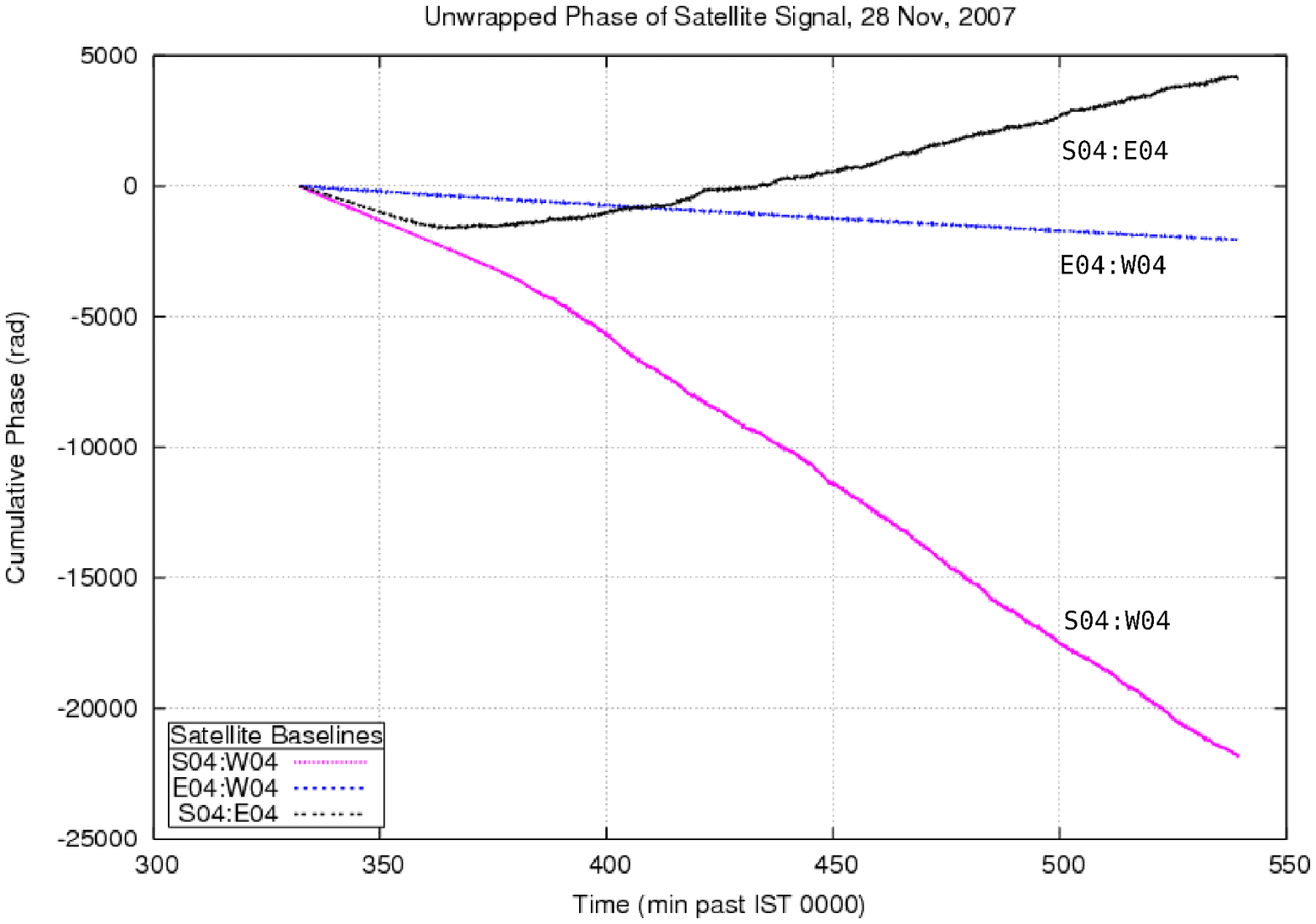}
\caption{Satellite cumulative phase seen varying smoothly over ~3hrs due to smooth diurnal sinusoidal motion of the satellite in its inclined orbit. The random jitter can be attributed to net phase distortions in the line of sight.}
\end{figure}

\section {Visibility Visualization and Data Quality Analyzer}

During our trial observations, the data were recorded over 16 MHz
bandwidth providing a 62.5 KHz spectral resolution and an integration
time of  131.072 ms, resulting in a data rate of ~50 GB/Hr. 
Consequently, we felt the need to develop new tools for visualization
and preprocessing of visibilities with particular attention to managing
large volumes of data.  Basic operations supported by our tool include
display of visibilities and processed outputs as images, reformatting to
various file formats, extraction/ reordering and recording of subsets
etc., with a baseline being the operational unit. A unified callback
interface exists which simplifies the plugging-in of modules for different
operations like correlation between different days, FFT along any axis
etc.  In our implementation, any data selected by the user is extracted
by a file reader object, pooled by a data formatter object and processed
by a centralized processor  object, which interfaces with a per baseline
display object. A central controller manages the graphical user interface
(GUI)  and sequences all the desired operations.

Fig. 2 gives a snapshot of our tool GUI in action. A feel for the capability of managing large volumes of data by our tool
can be obtained by noting that each frame in Fig. 2 corresponds to about
a Megabyte of visibility data and our tool can refresh a dozen such
frames simultaneously at 100 Hz.  Each horizontal panel shows interday
visibility cross-correlations and includes several frames of 256 x 400
pixels. In a given frame, while columns correspond to the 256 spectral
channels (16 MHz), each row corresponds to a time slice (0.131s).

\begin{figure}
\plotone{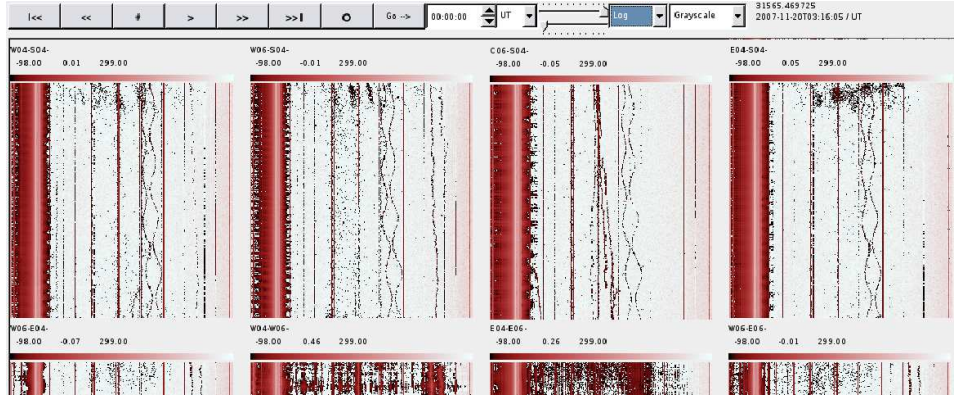}
\caption{Per channel (X axis) visibility cross-correlations over identical LST ranges on two days. Regions of high brightness over time (Y axis) indicate consistent data. Each frame in the panel represents the indicated baseline.}
\end{figure}

A major new feature of our tool for data editing pertains to correlations
of visibilities from multiple sessions with identical antenna settings
and LST ranges.  Fig. 2 shows such a correlation, in which regions of poor
correlation can be seen to be darker in color and correspond to local effects
like interference which will be uncorrelated from one day to the other
while the sky contribution should be identical.

An indication of the nature of RFI at GMRT is given by our analysis
of 8-second blocks of visibilities, which correspond to typical GMRT 
integration times.  In almost 70-80\% of such 8-second blocks, we found
 that $6-10\sigma$ deviations occurred about 10-20\% of the time, not 
necessarily contiguously within the block. Such deviations remain undetected 
by conventional flagging on visibilities integrated over 8 seconds or more.  
In contrast, they introduce systematic errors in the form of biases in the 
mean visibility which can seriously affect the quality of images.

\section {Conclusion}

A new tool has been developed for visualization and efficient analysis of
high volume interferometric datasets, while providing users with different
descriptions of errors in the dataset. These can provide better insights
towards the quality of data and help form a more reliable dataset to be
presented to regular post-processing and imaging software.  Support for
transit observations with the GMRT are discussed, which is expected to provide an
efficient observational mode for a rapid, low frequency survey.

\acknowledgements %%% Text of acknowledgements runs on after this command.

We thank the Director, NCRA and the Chief Scientist, GMRT for
the special allocation of test time for these observations. We gratefully
acknowledge the help of colleagues from GMRT. The software modifications
and the planning of the observational strategy would have been very
difficult for us without the help of Jayaram Chengalur from NCRA.
GMRT is run by the National Center for Radio Astrophysics of the Tata
Institute of Fundamental Research.

%%% THE BIBLIOGRAPHY
%%%
%%% CONSULT SECTION 3 OF "INSTRUCTIONS FOR AUTHORS" FOR HOW TO USE NATBIB.
%%% AUTHORS ARE ENCOURAGED TO USE EITHER THE "THEBIBLIOGRAPY" ENVIRONMENT
%%% BY UNCOMMENTING (DELETING THE "%" SYMBOL) THE COMMANDS BELOW, OR BY
%%% USING THE BIBTEX ENVIRONMENT. TO FIND OUT WHICH IS APPLICABLE TO YOUR
%%% CONTRIBUTION, CONSULT THE VOLUME EDITORS FOR YOUR PROCEEDINGS.
%%%

\end{document}